\documentclass[%
reprint,
amsmath,amssymb,
aps,
]{revtex4-1}

\usepackage{graphicx}% Include figure files
\usepackage{dcolumn}% Align table columns on decimal point
\usepackage{bm}% bold math
\usepackage{mathrsfs}
\begin{document}
	
	\preprint{APS/123-QED}
	
	\title{$\Omega$-deuteron Interaction in Folding Model}% Force line breaks with \\
	%\thanks{A footnote to the article title}%
	
	\author{Faisal Etminan }
	%\altaffiliation[Also at ]{Physics Department, XYZ University.}%Lines break automatically or can be forced with \\
	\affiliation{%
		Department of Physics, Faculty of Sciences, University of Birjand, Birjand 97175-615, Iran
	}%
	\email{fetminan@birjand.ac.ir}
	
	\author{Mohammad Mehdi Firoozabadi}%
	\email{mfiroozabadi@birjand.ac.ir}
	\affiliation{
		Department of Physics, Faculty of Sciences, University of Birjand, Birjand 97175-615, Iran
	}%
	
	%\collaboration{MUSO Collaboration}%\noaffiliation
	
	\date{\today}% It is always \today, today,
	%  but any date may be explicitly specified
	
	\begin{abstract}
		A simple single folding model for $\Omega$-deuteron with maximal spin $\left(I\right)J^{P}=\left(0\right)5/2^{+}$ is investigated. $\Omega$
		is assumed to orbit an unperturbed deuteron in a $\Omega$-deuteron
		potential based on a separable $\Omega$-nucleon potential from lattice QCD. We show that the effective central folding potential of $\Omega d$ in the $^{5}S_{2}$ channel has a simple Wood-Saxon form, and approximate the upper bound for the binding energy of $\Omega$ particle on a deuteron. In order to investigate how changes in the wave functions affect the results, we consider four analytical forms for $S$-state deuteron wave functions, i.e., two widely used Hulthén forms, as well as the modified Reid93 and Argonne v18 forms. Our calculations of binding energy from simple two-body approximation are compared with the results reported for the three-body problem; it is confirmed that the $\Omega d$ system is deeply bound. Although the single folding model reduces the three-body problem to a two-body problem, this simplification is inadequate.
	\end{abstract}
	
	\pacs{Valid PACS appear here}% PACS, the Physics and Astronomy
	% Classification Scheme.
	%\keywords{Suggested keywords}%Use showkeys class option if keyword
	%display desired
	\maketitle
	
	%\tableofcontents
	
	\section{Introduction} \label{sec:intro}
	A combination of simplicity and interesting interactions between nucleons and
	strange particles is a feature of $\Omega$ hypernuclei. Hypertriton \cite{cobis1997}, $\Xi d\:\left(I\right)J^{P}=\left(1/2\right)3/2^{+}$ \cite{Garcilazo2016} and $\Omega d $ in the state with maximal spin $\left(I\right)J^{P}=\left(0\right)5/2^{+}$ \cite{Garcilazo2018} have attracted considerable theoretical attention. 
	The interest is reflected in a number of investigations, where different techniques and models are used; for example, variational models \cite{kolesnikov1988},
	hyperspherical method \cite{verma1979}, Faddeev calculations \cite{miyagawa1993}, advanced few-body theoretical methods \cite{hiyama2019,Moosavi2019,Garcilazo2019}, lattice QCD calculations \cite{etminanmpla2014,nemura2014,Sasaki2018,sasaki2019} and simple $\Lambda$\textendash{}deuteron two-body models \cite{Congleton1992}.
	
	The first measurement of the
	proton-$\Omega$ correlation function \cite{Morita2016} in heavy-ion collisions by the STAR experiment \cite{Adam2019} at the Relativistic Heavy-Ion Collider (RHIC) favors the proton-$\Omega$ bound state hypothesis.
	
	Recently, an $\Omega d$ system in the state with maximal spin $\left(I\right)J^{P}=\left(0\right)5/2^{+}$ has been investigated by solving the three-body bound-state Faddeev equations; in these equations, the three components are assumed in S-wave state and the latest HAL QCD Collaboration for $\Omega N$ and $\Omega\Omega$ interactions. The results show that the $\Omega d$ system is bound with a binding energy of about 21 MeV \cite{Garcilazo2018,Garcilazo2019}. In the two-body approximation, scattering can be studied and compared with the full three-body results. Additionally, an effective two-body model might sometimes be useful in qualitative understanding and in preliminary estimates. We can perhaps get some insight by specifying that the folding approximation basically assumes a specific form for the total three-body wave function.
	
	This paper presents a simple model of the interaction between minus $\Omega$ 
	(with three strange quarks) and deuteron. The simple model is easy to apply and is considered as a reference point against which more sophisticated descriptions can be compared. The $\Omega d$ system is assumed to be a deuteron and a $\Omega$ particle moving in an effective $\Omega d$ potential. The nucleon part of the wave function is exactly that of a free deuteron and the Omega part is obtained from a $\Omega d$ potential constructed as follows. First, a separable fit is made to the $\Omega N$ s-wave potential, which is obtained from lattice QCD calculation by HAL QCD group \cite{Iritani2019prb,etminan2014}.
	Then, the $\Omega N$ potential is summed over the two nucleons and averaged over the deuteron wave function (DWF). Finally the resulting $\Omega d$ potential is fitted to a Wood-Saxon form, and only the s-wave part is
	used. Next, we solve the Schr\"{o}dinger equation with the fitted potential
	in the infinite volume and extract the binding energies. The model is expected to be accurate only for low energy properties of $\Omega d$ since it is based on a $\Omega N$ potential, which is fitted to the low energy $\Omega N$ scattering parameters.
	
	This paper is organized as follows. In Section ~\ref{sec:Spin-2--potential}, a brief description of the HAL QCD $\Omega N$ two-body interactions is presented. Section~\ref{sec:Two-body-approximations-to} presents a simple model that reduces a three-body problem to an effective two-body problem in order to study the interaction between $\Omega$ and deuteron. In Section~\ref{result}, the numerical results are discussed, and in particular, the two-body binding energy is compared to the three-body one. Finally, in Section~\ref{sec:Summary-and-conclusions}, summary and conclusion are presented.
	
	\section{Spin-2 $\Omega N$ potential\label{sec:Spin-2--potential}}
	S-wave and spin $2$ $\Omega N$ potential in configuration space
	is given by HAL QCD Collaboration with nearly physical quark masses 
	\cite{Iritani2019prb}. The discrete lattice potential is fitted by an analytic function composed of an attractive Gaussian core plus a long range
	$\mathrm{(Yukawa)^{2}}$ attraction with the form factor \cite{etminan2014}
	
	\begin{equation}
	V_{\Omega N}\left(r\right)=b_{1}e^{\left(-b_{2}r^{2}\right)}+b_{3}\left(1-e^{-b_{4}r^{2}}\right)\left(\frac{e^{-m_{\pi}r}}{r}\right)^{2},\label{eq:NOmega_pot}
	\end{equation}
	
	The pion mass in Eq.~\ref{eq:NOmega_pot} is taken from the simulation,
	$m_{\pi}=146$ MeV. The lattice results are fitted reasonably well,
	$\chi^{2}/d.o.f\simeq1$, with four different sets of parameters given
	in Table~\ref{tab:Fit_para}. 
	
	The low-energy observables for this potential are: a scattering
	length of $a_{0}^{\Omega N}=5.30$ fm, an effective range of $r_{eff}^{\Omega N}=1.26$ fm and a binding energy of $B^{\Omega N}=1.54\: $ MeV \cite{Iritani2019prb}.
	
	\begin{table}[h]
		\caption{Fitting parameters in Eq.~\ref{eq:NOmega_pot} for different models,
			$P_{i}$ , of $^{5}S_{2}$ $\Omega N$ interaction \cite{Iritani2019prb}.\label{tab:Fit_para}}
		\begin{ruledtabular}
			\begin{tabular}{ccccc}
				& $P_{1}$ & $P_{2}$ & $P_{3}$ & $P_{4}$\tabularnewline
				\hline  
				$b_{1}\left(\mathrm{MeV}
				\right)$ & -306.5  & -313.0 & -316.7 & -296\tabularnewline
				$b_{2}\left(fm^{-2}\right)$ & 73.9  & 81.7 & 81.9  & 64\tabularnewline
				$b_{3}\left(\mathrm{MeV}.fm^{-2}\right)$ & -266  &  -252 & -237 & -272\tabularnewline
				$b_{4}\left(fm^{-2}\right)$ & 0.78  & 0.85  & 0.91  & 0.76\tabularnewline
			\end{tabular}
		\end{ruledtabular}
	\end{table}
	
	\section{Two-body approximations to the three-body problem\label{sec:Two-body-approximations-to}}
	Reduction of a three-body problem to an effective two-body one has computational advantages. It is tempting to make such a reduction for $\Omega d$, which, approximately, can be considered as $\Omega$-particle and deuteron. This case appears to be very well suited for a two-body description, as previously attempted \cite{Congleton1992,etminan2019}.
	The first step of the procedure is obviously to construct the $\Omega$\textendash{}deuteron effective potential by folding the deuteron wave function with the $\Omega N$ interaction. The remaining steps are then of two-body nature, and therefore, computationally much simpler. 
	
	The interaction between the $\Omega$-particle and deuteron is given by the Watanabe ansatz, and the deuteron optical potential $U_{F,\Omega d}\left(\mathbf{R}\right)$ is given by \cite{Watanabe1958}
	
	\begin{eqnarray}
	&& U_{F,\Omega d}\left(\mathbf{R}\right)=  {\textstyle }\\
	&& \int\mathbf{r}_{d}\Psi_{d}^{*}\left(\mathbf{r}_{d}\right)\left[V_{\Omega p}\left(\mathbf{R}+\frac{1}{2}\mathbf{r}_{d}\right)+V_{\Omega n}\left(\mathbf{R}-\frac{1}{2}\mathbf{r}_{d}\right)\right]\Psi_{d}\left(\mathbf{r}_{d}\right),\nonumber 
	\end{eqnarray}
	where $V_{\Omega p}$ and $V_{\Omega n}$ denote two-body spin-independent central potentials of the $\Omega-$proton and $\Omega-$neutron systems, respectively; these are calculated at half the deuteron energy. Note that the integration runs over the internal coordinates $\mathbf{r}_{d}$ of the deuteron. For the central potential terms, this gives a very good approximation 
	\cite{Keaton1973},
	
	\begin{eqnarray}
	&& U_{F,\Omega d}^{c}\left(\mathbf{R}\right)  =\label{eq:vc_fold}\\
	&&  \int d\mathbf{r}_{d}\rho_{d}\left(\mathbf{r}_{d}\right)\left[V_{\Omega p}^{c}\left(\mathbf{R}+\frac{1}{2}\mathbf{r}_{d}\right)+V_{\Omega n}^{c}\left(\mathbf{R}-\frac{1}{2}\mathbf{r}_{d}\right)\right],\nonumber 
	\end{eqnarray}
	where $\rho_{d}=\Psi_{d}^{*}\Psi_{d}$ denotes the deuteron density and the index $c$ represents central potential terms.
	
	Deuteron wave functions are written as the sum of the wave functions for
	$^{3}S_{1}$- and $^{3}D_{1}$-state ~\cite{Blatt}
	
	\begin{equation}
	\Psi_{d}=\psi_{S}+\psi_{D}=\frac{u\left(r\right)}{r}\mathscr{Y}_{101}^{1}+\frac{w\left(r\right)}{r} \mathscr{Y}_{121}^{1},
	\end{equation}
	where $u\left(r\right)$ and $w\left(r\right)$ are the radial deuteron wave functions for states with orbital moments of $l=0$ and $2$;
	$\mathscr{Y}_{JlS}^{M}$ are the normalized spin angle wave functions, belonging to a state of total angular momentum $J$ whose $z$ component is $M$. Actually, $J$ is the combination of the spin ($S$) and orbital($l$)  angular momenta. For a deuteron $J=M=S=1$. The condition of normalization for DWF $\Psi_{d}$ can be written as 
	\begin{equation}
	p_{S}+p_{D}={\textstyle \int_{0}^{\infty}\left(u^{2}\left(r\right)+w^{2}\left(r\right)\right)dr=1},
	\end{equation}
	where $p_{S}$ and $p_{D}$ are probabilities to find a deuteron in $S$- and $D$-state, respectively. A rough estimate can be obtained by neglecting the small probability of $D$-state compared to unity and only consider the deuteron $S$-state probability~\cite{Blatt}.

	In order to investigate how changes in the wave function affect the results, we consider four type forms for $S$-state DWF, two analytical widely used Hulthén forms (i.e., $u_{H}^{(1)}(r)$  and $u_{H}^{(2)}(r)$), as well as the modified analytical forms of DWFs from Reid93 and Argonne v18 potentials~\cite{Zhaba2016}.
	
	Historically, the analytical ad hoc Hulthén function has been so often used in calculations involving the $S$-state DWF and deuteron matrix elements that it would be hopeless to give a list of references. Here, Type 1 Hulthén (Hulthén-1) form is given by ~\cite{adler1977}
	\begin{equation}
	u_{H}^{\left(1\right)}(r)=N\left(e^{-\alpha^{\prime}r}-e^{-\beta^{\prime}r}\right)/r, \label{eq:hul1}
	\end{equation}
	where $N^{2}=0.783$, $\alpha^{\prime}=0.2316\:fm^{-1}$ and $ \beta^{\prime}=5.98\:\alpha^{\prime}$. The normalization constant $N$ and two parameters of the Hulthén-1 function are determined by the deuteron binding energy and the triplet effective range. Type 2 Hulthén (Hulthén-2) form is given by~\cite{MROWCZYNSKI1992}
	
	\begin{equation}
	u_{H}^{\left(2\right)}(r)=\sqrt{\frac{\alpha\beta\left(\alpha+\beta\right)}{2\pi\left(\alpha-\beta\right)^{2}}}\frac{e^{-\alpha r}-e^{-\beta r}}{r}, \label{eq:hul2}
	\end{equation}
	where $ \alpha=0.23\: fm^{-1} ; \beta=1.61\: fm^{-1} $; these two parameters of Hulthén-2 function are determined by the deuteron binding energy and the singlet effective range. For many potential models, the wave functions coincide very closely with the simple Hulthén function, except, of course, in the region $r\sim0.5$ fm, where many potentials have a hard core.
	
	Modified analytical forms of DWFs from Argonne v18 and Reid93 potentials can be parametrized in the following analytical expressions~\cite{Zhaba2016}:
	
	\begin{equation}
	\begin{cases}
	u \left(r\right) & =r^{3/2}\sum_{i=1}^{N}A_{i}\exp\left(-a_{i}r^{3}\right),\\
	w \left(r\right) & =r\sum_{i=1}^{N}B_{i}\exp\left(-b_{i}r^{3}\right). \label{eq:av18}
	\end{cases}
	\end{equation}
	
	The coefficients of new analytical forms for DWFs in coordinate space for  Reid93 and Argonne v18 potentials have been numerically calculated and given, respectively, in Tables 2 and 3 of Ref.~\cite{Zhaba2016}.
		
		The Coulomb potential increases the binding for systems containing a proton compared to those with a neutron. This is due to the attractive $\Omega^{-}p$ interaction. We included the Coulomb interaction as
		\begin{equation}
		V_{C}\left(r\right)=\pm\alpha_{f}\frac{e^{-r/r_{0}}}{r},\label{eq:V-coulomb}
		\end{equation}
		here $\alpha_{f}$ is the fine structure constant and $r_{0}$ is the screening radius, which is taken to be $r_{0}=50$ fm.
		
		\section{Results} \label{result}
		
\begin{figure}
			\includegraphics[scale=0.72]{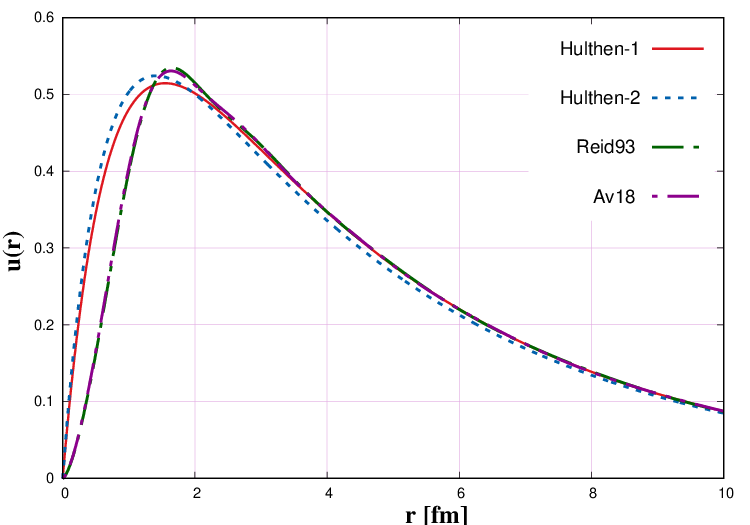}
			\caption{S-state forms of DWF for four types are presented for comparison. Here, Hulthén-$1$ is $u_{H}^{(1)}(r)$ (Eq.~\ref{eq:hul1}), Hulthén-$2$ is $u_{H}^{(2)}(r)$ (Eq.~\ref{eq:hul2}) and Reid93 and Argonne v18 are the modified analytical forms of DWFs from Argonne v18 and Reid93 potentials (Eq.~\ref{eq:av18}) given in Ref.~\cite{Zhaba2016}.     \label{fig:dwf}}
		\end{figure}

		S-state forms of DWFs for four types, that is, $u_{H}^{(1)}(r)$ (Hulthén-$1$) in Eq.~\ref{eq:hul1}, $u_{H}^{(2)}(r)$ (Hulthén-$2$) in Eq.~\ref{eq:hul2} and Reid93 and Argonne v18 in Eq.~\ref{eq:av18}, are shown in Fig.~\ref{fig:dwf}.
		
		To obtain observables such as the scattering phase shift and the binding energy we fit the resulting $\Omega d$ potential by a Woods-Saxon type function form
		
		\begin{equation}
		V_{\Omega d}^{fit}(r)=-V_{0}\left[1+\exp\left(\frac{r-R}{c}\right)\right]^{-1}. \label{eq:Vfit Omegad}
		\end{equation}
		
\begin{figure}
			\includegraphics[scale=0.72]{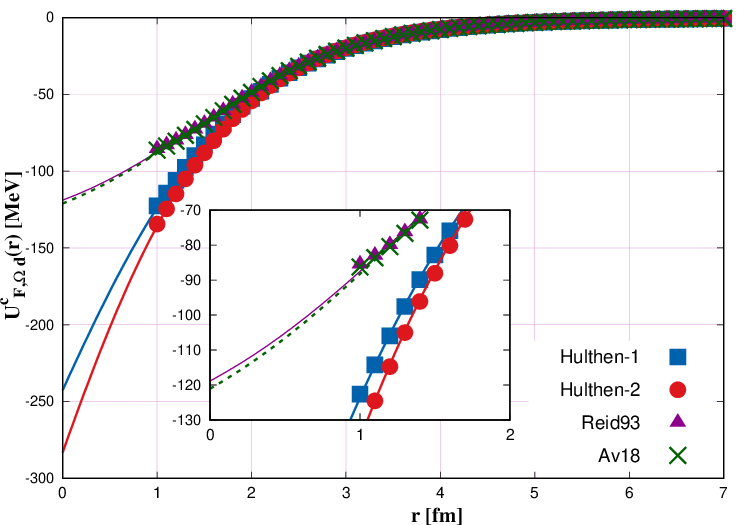}
			\caption{Effective central folding potential $U_{F,\Omega d}^{C}$ in the $^{5}S_{2}$ and the result of fitting (solid lines) by using $V_{\Omega d}^{fit}(r)$ in Eq.~\ref{eq:Vfit Omegad}, which are presented for four types of  DWFs, i.e., Hulthén-$1$, Hulthén-$2$, Reid93 and Argonne v18.\label{fig:The-effective-central}}
		\end{figure}

		The results of the fit and the corresponding parameters, $ V_{0}, R $ and $c$, are shown in Fig.~\ref{fig:The-effective-central} and Table~\ref{tab:The-fitting-parameters-Omega-d}.
		Then, we solve the Schr\"{o}dinger equation with the fitted potential in the infinite volume to extract its binding energy.
		
		\begin{table}[h]
			\caption{The fitting parameters in Eq.~\ref{eq:Vfit Omegad} (in physical units) and the corresponding two-body binding energies, $B_{\Omega d}^{2b}$ (MeV) of the $\left(I\right)J^{P}=\left(0\right)5/2^{+}$ $\Omega d$ state with respect to the $\Omega NN$ threshold for $P_{1}$ model of $\Omega N$ interaction given in Table~\ref{tab:Fit_para} \cite{Iritani2019prb}. The results have been obtained using the experimental masses of $N$ and $\Omega$ that are equal to $938.9\: \mathrm{MeV}/c^{2}$ and $1672.45\:\mathrm{MeV}/c^{2}$, respectively. The results corresponding to the $N$ and $\Omega$ masses are, respectively, $954.7\: \mathrm{MeV}/c^{2}$ and $1711.5\: \mathrm{MeV}/c^{2}$, which have been derived by the HAL QCD Collaboration and are indicated in parentheses \cite{Gongyo2018,Iritani2019prb}. The binding energy from three-body calculation in Ref.~\cite{Garcilazo2019} is $20.9 (22.0)$ MeV, which has been measured with respect to the $\Omega NN$ threshold.  \label{tab:The-fitting-parameters-Omega-d}}
			\begin{ruledtabular}
				\begin{tabular}{ccccccc}		
					Model& $V_{0} $ (MeV)& $ R $(fm)& $ c $(fm)&$B_{\Omega D}$(MeV) \tabularnewline
					\hline 	
					Hulthén-1  & 538 & 0.98 & -0.19 & 15.7(16.4) \tabularnewline
					Hulthén-2  & 675 & 0.94 & -0.31 & 19.2(20.0) \tabularnewline
					Reid93     & 143 & 0.89 &  1.39 &  6.7(7.0) \tabularnewline
					Av18       & 148 & 0.90 &  1.34 &  6.8(7.2) \tabularnewline
				\end{tabular}
			\end{ruledtabular}
		\end{table}
		
		In Table~\ref{tab:The-fitting-parameters-Omega-d}, we show the binding energies obtained from folding potentials of the state with maximal spin, $\left(I\right)J^{P}=\left(0\right)5/2^{+}$. These results have been obtained for different types of DWFs for $P_{1}$ model of $\Omega N$ interaction  reported in Ref.~\cite{Iritani2019prb} and summarized in Table~\ref{tab:Fit_para}. The binding energies have been measured with respect to the $\Omega NN$ threshold. The binding energy, $B_{\Omega d}^{3b}$, from three-body calculations in Ref.~\cite{Garcilazo2019} is $20.9 (22.0) $ MeV, which has been measured with respect to the $\Omega NN$ threshold (indicated in parentheses). The results corresponding to the $N$ and $\Omega$ derived by HAL QCD Collaboration are $954.7\: \mathrm{MeV}
		/c^{2}$ and $1711.5\: \mathrm{MeV}/c^{2}$, respectively \cite{Gongyo2018,Iritani2019prb}). One can see from Table~\ref{tab:The-fitting-parameters-Omega-d} that the binding energies calculated by Hulthén-1 ($ 15.7 $ MeV) and Hulthén-2 ($ 19.2 $ MeV) DWFs are more closer to the binding energy from three-body calculations. Since the real potential must be more attractive than the folding potential, the resulting energy can only be an upper bound for $\Omega d$ systems. The almost big difference between the resultant binding energies from Hulthén forms and  Reid93/Av18 potentials can originate from the behavior of the corresponding wave functions in the range $ r\sim 0-1 $ fm. As can be seen from Fig.~\ref{fig:dwf}, in this range, the Hulthén form wave functions are more stronger than Reid93 and Av18 wave functions.    
		
		\section{Summary and conclusions\label{sec:Summary-and-conclusions}}
		It has been recently shown in lattice QCD analysis by HAL QCD Collaboration that $\Omega N$ interacting potentials have nearly physical quark masses ($m_{\pi}\simeq146$ MeV and $m_{K}\simeq525$ MeV). This analysis showed an attractive potential in the $\Omega N$ $^{5}S_{2}$ channel, which supports a bound state with a central binding energy of $1.54$ MeV. 
		
		In turn, the $\Omega NN$ and $\Omega\Omega N$ three-body systems have been examined in \cite{Garcilazo2018,Garcilazo2019} by using the latest HAL QCD Collaboration $\Omega N$ and $\Omega\Omega$ interactions. These results have shown that the $\Omega d$ system in the state with maximal spin $\left(I\right)J^{P}=\left(0\right)5/2^{+}$ is bound with a binding energy of about $21$ MeV. 
		
		We tested the two-body approximation obtained by the DWF folding and $\Omega$\textendash{}nucleon interaction. In order to investigate how changes in the wave function affect the results, we computed binding energies for four type form $S$-state DWFs, i.e., two widely used analytical Hulthén forms, as well as modified analytical forms of DWFs from Reid93 and Argonne v18 potentials. 
		
		We showed that the effective central folding potential of $\Omega d$ in the $^{5}S_{2}$ channel has a simple Wood-Saxon form. We calculated the binding energy for simple two-body approximation by Hulthén forms and found that the results are closer to the results reported for the three-body problem \cite{Garcilazo2019}. Although the approximation in the present investigation is inadequate for better details and accuracy, an effective two-body model can sometimes be useful for qualitative understanding and preliminary estimates. The folding approximation basically assumes a specific form for the total three-body wave function. Since the potential must be more attractive than the folding potential, the resulting energy is only an upper bound for the $\Omega d$ systems. 
		
		Our results confirmed that the $\Omega d $ systems are deeply bound states or resonances that may be found experimentally in the real world. In Ref. \cite{Morita2016}, it is discussed that how the two-particle momentum correlation between a proton and a $\Omega$ baryon in high-energy heavy ion collisions (e.g., STAR experiment at RHIC \cite{Adam2019}) may unveil the existence of these states.
		
		We are going to carry out three-body scattering problem in a future work. The strict three-body calculations of scattering length and effective range (where a $\Omega$-particle is scattered on a deuteron) are necessary to obtain detailed information about the $\Omega d$ system structure or the two-body interactions describing this three-body system. The scattering length is closely related to the size of the binding energy, and the effective range is of the order of the range of the effective three-body radial potential.
		% The \nocite command causes all entries in a bibliography to be printed out
		% whether or not they are actually referenced in the text. This is appropriate
		% for the sample file to show the different styles of references, but authors
		% most likely will not want to use it.
		\nocite{*}
		
		\bibliography{dOmega_ver03}% Produces the bibliography via BibTeX.
		
	\end{document}